\documentclass[11pt]{article}
\usepackage{amsmath}
\usepackage{amssymb}
\usepackage[dvipdfmx]{graphicx}
\usepackage{bm}
\usepackage{latexsym}
\usepackage{color}

\newcommand{\ltsim}{\protect\raisebox{-0.7ex}{$\:\stackrel{\textstyle <}{\sim}\:$}} 
\newcommand{\gtsim}{\protect\raisebox{-0.7ex}{$\:\stackrel{\textstyle >}{\sim}\:$}} 

\numberwithin{equation}{section}

\begin{document}

\begin{titlepage}

\setcounter{page}{1} \baselineskip=15.5pt \thispagestyle{empty}

\begin{flushright}
DESY 13-169, RUP-13-10, RESCEU-4/13
\end{flushright}
\vfil

\bigskip
\begin{center}
{\LARGE \textbf{Graceful exit from Higgs G-inflation}}
\vskip 15pt
\end{center}

\vspace{0.5cm}
\begin{center}
{\Large 
Kohei~Kamada$^{a,}$\footnote{kohei.kamada@desy.de},
Tsutomu~Kobayashi$^{b,}$\footnote{tsutomu@rikkyo.ac.jp},
Taro~Kunimitsu$^{c,d,}$\footnote{kunimitsu@resceu.s.u-tokyo.ac.jp},
Masahide~Yamaguchi$^{e,}$\footnote{gucci@phys.titech.ac.jp},
and Jun'ichi~Yokoyama$^{d,f,}$\footnote{yokoyama@resceu.s.u-tokyo.ac.jp}
}
\end{center}

\vspace{0.3cm}

\begin{center}
\textit{$^{a}$ Deutsches Elektronen-Synchrotron DESY, Notkestrasse 85, D-22607 Hamburg, Germany}\\

%\vskip 4pt
%\textit{
%  Institut de Th\'eorie des Ph\'enom\`enes Physiques,
% \'Ecole Polytechnique F\'ed\'erale de Lausanne,
%  CH-1015 Lausanne, Switzerland}

\vskip 4pt
\textit{$^{b}$ Department of Physics, Rikkyo University, Toshima, Tokyo 175-8501, Japan}\\

\vskip 4pt
\textit{$^{c}$ Department of Physics, Graduate School of Science,\\
The University of Tokyo, Tokyo 113-0033, Japan}\\

\vskip 4pt
\textit{$^{d}$ Research Center for the Early Universe (RESCEU), Graduate
School of Science, \\The University of Tokyo, Tokyo 113-0033, Japan}\\

\vskip 4pt
\textit{$^{e}$ Department of Physics, Tokyo Institute of Technology, Tokyo 152-8551, Japan}\\

\vskip 4pt
\textit{$^{f}$ Kavli Institute for the Physics and Mathematics of the Universe
(Kavli IPMU), \\WPI, TODIAS, The University of Tokyo, Kashiwa, Chiba, 277-8568, Japan}
\end{center} \vfil

\vspace{0.8cm}

\noindent 
Higgs G-inflation is a Higgs inflation model with a generalized Galileon
term added to the standard model Higgs field, which realizes inflation
compatible with observations.  Recently, it was claimed that the
generalized Galileon term induces instabilities during the oscillation
phase, and that the simplest Higgs G-inflation model inevitably suffers from this problem.
In this paper, we extend the original Higgs G-inflation
Lagrangian to a more general form, namely introducing a higher-order
kinetic term and generalizing the form of the Galileon term, so that the
Higgs field can oscillate after inflation without encountering
instabilities.  
Moreover, it accommodates a large region of the $n_s$ - $r$
plane, most of which is consistent with current observations, 
leading us to expect the detection of B-mode polarization in the cosmic microwave background in the near future.
\vfil

\end{titlepage}

\newpage
%\setcounter{tocdepth}{2}
%\tableofcontents
\setcounter{footnote}{0}

\section{Introduction}
\label{sec:intro}

Cosmic inflation \cite{Starobinsky:1980te, Sato:1980yn, Guth:1980zm},
which is now considered to be an indispensable part of standard
cosmology, requires an effective scalar field in order to keep the
universe in a quasi-de Sitter phase. Such scalar fields are thought to
be ubiquitous in theories beyond the standard model (SM) of particle physics,
but the only scalar field identified so far is the 125 GeV scalar boson %Higgs field 
recently discovered at the LHC \cite{atlas:2012, cms:2012}, whose
characteristics have up to date no deviation from the predictions of the
SM Higgs. %standard model.

Although there are many indications of physics beyond the SM, 
%standard model, there has been 
no direct evidence has been reported until now %so far 
except for neutrino oscillations \cite{Fukuda:1998mi, GonzalezGarcia:2007ib}. This motivates us to consider the possibility that the
Higgs field, the only scalar field in the SM, is responsible
for inflation, assuming that the discovered particle is the SM Higgs.
With no hints of physics beyond the SM at the
LHC, this type of inflation models has to be taken seriously.

It is known that inflation driven by the SM Higgs field,
with the canonical kinetic term and renormalizable potential,
does not reproduce the observable universe, since both the curvature
and tensor perturbations produced are much larger than what we observe today.  
Consequently, there
have been various proposals for Higgs inflation models by extending the
structure of the SM Lagrangian, starting with the inclusion
of a large non-minimal coupling term with gravity
\cite{CervantesCota:1995tz, Bezrukov:2007ep, Barvinsky:2008ia}. Other
models include a derivative coupling with the Einstein tensor
(new Higgs inflation) \cite{Germani:2010gm}, a galileon-like term (Higgs
G-inflation) \cite{Kamada:2010qe}, and a non-canonical kinetic term
(running kinetic inflation) \cite{Nakayama:2010kt}, all of which can be
treated in a unified manner in the context of generalized G-inflation
\cite{Kobayashi:2011nu} as generalized Higgs inflation
\cite{Kamada:2012se}\footnote{Another Higgs inflation model without any 
higher derivative couplings was also proposed recently \cite{Hamada:2013mya}.}.

Among these, Higgs G-inflation \cite{Kamada:2010qe} is distinct from the
other possibilities in the sense that it does not involve non-minimal
couplings between the Higgs field and gravity, and although it includes
higher derivative terms in the Lagrangian, it keeps the derivatives in
the equation of motion at second order. The original Higgs G-inflation
considered in Ref.~\cite{Kamada:2010qe} has an additional term in the Higgs
field Lagrangian of the form
\begin{equation} 
\frac{\varphi X}{M^4} \Box \varphi,
\end{equation}
which is a generalization of the Galileon term \cite{Deffayet:2010qz, Kobayashi:2010cm}, and is the simplest possibility of this type. Here,
$\varphi$ is the %neutral component of the 
Higgs field in the unitary gauge, $X$ is its
canonical kinetic function, $X:= -\partial_\mu\varphi\partial^\mu\varphi/2$ (we use the mostly plus sign convention for the metric),
and $M$ is some mass scale.

Recently, there was a claim that the Higgs field oscillation after
inflation does not occur for the above Galileon-inspired term
\cite{Ohashi:2012wf}. This is due to the large effect of the Galileon
term on the oscillation dynamics, which makes the coefficient of
$\ddot{\varphi}$ vanish in the equation of motion of the Higgs field,
thereby ending up with a singularity.

By taking a small value for the Higgs self-coupling constant $\lambda$,
oscillation can be realized in this model. Even in this case, however, the
sound speed squared of the curvature perturbation possibly becomes negative,
which makes the curvature perturbations obey a Laplacian equation
instead of a wave equation, leading to instabilities at small scales. As a 
result, a robust oscillation of the Higgs field occurs only if
\begin{equation}
\lambda<2.9\times10^{-9},
\end{equation}
which is far smaller than the value for the SM Higgs field.

In this paper, we extend the above model to a broader framework, and
propose a new Higgs G-inflation model in which the Higgs field
oscillates after inflation keeping $c_s^2>0$ even with phenomenologically 
natural values of the Higgs quartic coupling.  We will show that
by extending the structure of the kinetic sector in Higgs G-inflation
models, oscillation does occur in Higgs G-inflation models while keeping
$c_s^2$ positive. We will also present a convenient method 
of calculating the predictions of the generalized model.
We concentrate on inflation realized by the
Higgs field, but the analysis here can easily be extended to general
potential driven G-inflation models.

Before we go on, we would like to note the value of the Higgs 
quartic coupling constant $\lambda$.  For the 125 GeV SM Higgs field,
the value of $\lambda$ at the electroweak scale is determined by 
the mass of the Higgs boson, which gives
\begin{equation}
\lambda = \frac{m_h^2}{2v^2}\simeq 0.13.
\end{equation}
When considering inflation, we need to take into account the running of
the coupling constant. Using the renormalization group equations for the
SM \cite{Casas:1994qy, EliasMiro:2011aa}, and assuming that
the Galileon term does not change the equations substantially, $\lambda$
becomes logarithmically smaller at higher energy
scales.\footnote{One may wonder if $\lambda$ can be (accidentally)
of the order of $10^{-13}$ at the inflationary scale, which explains the
correct magnitude of the primordial curvature perturbations with
robust oscillation after inflation. However, even in this case, the
tensor-to-scalar ratio is too large to be compatible with the Planck
results as long as the logarithmic correction is not so significant.}
The value could become negative, depending on the values of the top
quark mass and the strong coupling constant, but once the Higgs field acquires
an expectation value 
where $\lambda$ is negative, inflation never occurs and it cannot fall down to the electroweak vacuum. 
%the standard vacuum state cannot be realized. 
Thus, we assume here that the Higgs field was in an initial
condition at which it can drive inflation with positive $\lambda$. At this scale, the value of
$\lambda$ would be $\mathcal{O}(0.01)$ assuming no fine-tuning. For this
reason, we will take $\lambda$ to be 0.01 in the numerical analyses of
this paper, although the results of this paper would not strongly depend on the
precise value.

The paper is organized as follows. In the next section, we review the
original Higgs G-inflation model, and show where the problem lies. In
Sec. \ref{sec3} we extend the form of the Lagrangian in order to
realize inflation and subsequent field oscillation consistently, 
and present the results from numerical
calculations. The final section is devoted to
conclusions and discussions. In the Appendixes, we summarize the basic
formulas for the equations of motion and the primordial perturbations.

%%%%%%%%%%%%%%%%%%%%%%%%%%%%%%%%%%%%%%%%%%%%
%%%%%%%%%%%%%%%%%%%%%%%%%%%%%%%%%%%%%%%%%%%%

\section{Higgs G-inflation and its reheating phase}
\label{sec2}

In this section, we review the Higgs G-inflation model and its possible
instabilities during the reheating phase.

The SM Higgs Lagrangian is written as
\begin{equation}
S= \int d^4 x \sqrt{-g}\left[ \frac{M_{P}^2}{2}R -|D_\mu \mathcal{H}|^2
			- \lambda\left(|\mathcal{H}|^2 - v^2\right)^2 \right],
\end{equation}
where $M_{P}$ is the reduced Planck mass, $R$ is the Ricci scalar, and
$\mathcal{H}$ is the Higgs doublet. Higgs G-inflation
is driven by the neutral component of the SM Higgs
field, or the scalar Higgs boson in the unitary gauge, so we will 
focus on this component. The action for the SM Higgs field becomes of the form
\begin{equation}
S= \int d^4 x \sqrt{-g}\left[ \frac{M_{P}^2}{2}R + X
					-V(\varphi) \right], 
\end{equation}
where $\varphi$ is the neutral component of the Higgs field.
Here and hereafter we omit
interactions with massive gauge fields that are irrelevant to the dynamics of
inflation. We consider the case where the neutral component of the Higgs
field has a large value compared to the electroweak vacuum expectation
value, $v=246\ \mathrm{GeV}$. In this case, the Higgs potential can be
approximated as a quartic potential:
\begin{equation}
V(\varphi) \simeq \frac{1}{4}\lambda \varphi^4.
\end{equation}

In order to realize inflation consistent with observations, we need to
extend the structure of the Higgs field action. Higgs G-inflation
\cite{Kamada:2010qe} is a model in which an additional
self-interaction term of the form
\begin{equation}
- G(\varphi, X) \Box \varphi
\end{equation}
is introduced to the Higgs field Lagrangian. This term is a generalized Galileon term
\cite{Deffayet:2010qz, Kobayashi:2010cm}, which keeps the derivatives in
the equation of motion at second order. Adding this generalized Galileon
term, the full action becomes
\begin{equation}
S= \int d^4 x \sqrt{-g}\left[ \frac{M_{P}^2}{2}R + X -
			\frac{1}{4}\lambda \varphi^4 - G(\varphi, X)
			\Box \varphi \right]. 
\end{equation}
$G$ is an arbitrary function of $\varphi$ and $X$, except that it should 
not  contain any even powers of
%have an odd number of 
$\varphi$ in order to keep the gauge invariance of the action. In the original Higgs G-inflation model, 
\begin{equation}
G(\varphi, X) = -\frac{\varphi X}{M^4} \label{original}
\end{equation}
was adopted. $M$ here is a constant with a dimension of mass, which would be related to %signifies
the scale where new physics comes in\footnote{Here we assume that the new physics does 
not affect the inflation dynamics except for modification of the equation of motion due to the Galileon term.}. 
If $M$ is large enough compared to
the electroweak scale, this term would not affect the predictions in
collider experiments, and hence becomes compatible with the known
characteristics of the Higgs field.

The equation of motion for the Higgs field corresponds to
Eq.~(\ref{back13}) in Appendix B with
$m=1$, $n=0$, and $\widetilde{M} \rightarrow \infty$,
\begin{align}
0&=
 \left(1-6H\frac{\varphi\dot{\varphi}}{M^4}+2\frac{\dot{\varphi}^2}{M^4} 
 + \frac{3\varphi^2\dot{\varphi}^4}{2M_P^2M^8}\right)\ddot{\varphi} +3H\dot{\varphi} +\lambda \varphi^3 \notag\\
&\ \  -\left(9H^2-\frac{3\dot{\varphi}^2}{2M_P^2}-\frac{3\dot{\varphi}^4}{2M_P^2M^4}+\frac{3\lambda\varphi^4}{4M_P^2}\right) \frac{\varphi \dot{\varphi}^2}{2M^4} 
\label{HiggsGnum} \\ 
&=H\dot{\varphi}\left[3 - \eta - H\frac{ \varphi \dot{\varphi}}{M^{4}}
\left(9- 3 \epsilon  - 6\eta +2\eta\alpha \right) \right]+ \lambda
\varphi^3, 
\label{HiggsGslow} 
\end{align}
where $\epsilon$, $\eta$, and $\alpha$ are the slow-roll parameters
defined by
\begin{align}
\epsilon:=-\frac{\dot H}{H^2},
\quad
\eta:=-\frac{\ddot\varphi}{H\dot\varphi},
\quad
\alpha:=\frac{\dot \varphi}{H \varphi}.
\end{align}

The original Higgs G-Inflation model is realized when the Higgs field has a
large expectation value, where the slow-roll conditions
\begin{equation}
|\epsilon|,\ |\eta|,\ |\alpha| \ll 1,
\end{equation}
and the Galileon domination condition
\begin{equation}
|H\dot{\varphi}G_X| =\left|H\frac{ \varphi \dot{\varphi}}{M^4}\right|\gg
 1,
\end{equation}
are satisfied. Here the dynamics of the Higgs field is described by 
the slow-roll equation of motion,
\begin{equation} 
- 9H^2\frac{\varphi \dot{\varphi}^2}{M^4} + \lambda \varphi^3 \simeq 0. 
\end{equation}
$M$ is determined by the observed amplitude of the%value of the 
curvature perturbations. % produced.
The second order action for the comoving curvature perturbation $\zeta$
is given by \cite{Kobayashi:2011nu}
\begin{equation}
S_2 = M_{P}^2 \int d^4 x a^3\left[\mathcal{G}_S \dot{\zeta}^2
			      -\frac{\mathcal{F}_S}{a^2}(\vec{\nabla}\zeta)^2\right],
\end{equation}
where $\mathcal{F}_S$ and $\mathcal{G}_S$ are described by the background
quantities and are given in Eqs.~(\ref{pertFs}) and (\ref{pertGs}) for more general cases in
Appendix B. In the present case, from Eqs.~(\ref{pertFs}) and
(\ref{pertGs}) with $m=1$, $n=0$, and $\widetilde{M} \rightarrow
\infty$, we see that
\begin{align}
 \mathcal{F}_S &\simeq -\frac{2\varphi\dot{\varphi}^3}{HM_{Pl}^2M^4}, 
\\   
 \mathcal{G}_S &\simeq -\frac{3\varphi\dot{\varphi}^3}{HM_{Pl}^2M^4},
\end{align}
which yield the sound speed squared,
\begin{equation}
 c_s^2 := \frac{\mathcal{F}_S}{\mathcal{G}_S} \simeq \frac23.
\end{equation}
The power spectrum of the comoving curvature perturbation, $\zeta$, is
estimated in Eq.~(\ref{Gpower}) as
\begin{equation}
 \mathcal{P}_\zeta = \frac{1}{8\pi^2 c_s \mathcal{F}_S}
  \left(\frac{H}{M_P}\right)^2
  = -\frac{H^3M^4}{16\pi^2c_s \varphi \dot{\varphi}^3}.
\label{eqn:zetanorm}
\end{equation}
Using the Planck normalization, $\mathcal{P}_\zeta \simeq 2.2 \times
10^{-9}$ for $k=0.05\ \mathrm{Mpc}^{-1}$ yields
\begin{equation}
M \simeq 4.5 \times 10^{-6} \lambda^{-\frac{1}{4}}M_P \simeq 10^{13}
\; \mathrm{GeV}.
\end{equation}
The spectral index and tensor-to-scalar ratio are given as
\begin{equation}
n_s = 1-4\epsilon \simeq 0.967,
\end{equation}
\begin{equation}
r = \frac{64}{3}\left(\frac{2}{3}\right)^{\frac{1}{2}} \epsilon \simeq
 0.14,
\end{equation}
respectively, and there is a consistency relation between $r$ and the tensor spectral index $n_T$: 
\begin{equation}
r = -\frac{32\sqrt{6}}{9}n_T. \label{consistency}
\end{equation}

For the universe to transform into the hot big bang state after
inflation, the Higgs field has to oscillate around the minimum of the
potential and decay into the SM particles. But recently, it
was found that the Higgs field could not oscillate after inflation for
this model \cite{Ohashi:2012wf}.  The coefficient of $\ddot{\varphi}$ in
Eq.~(\ref{HiggsGnum}) vanishes before the Higgs field reaches the maximum
point after passing the potential minimum, where
$\ddot{\varphi}$ goes to infinity, leading to catastrophe.

Moreover, before reaching the point where $\ddot{\varphi}$ diverges,
 the value of the sound speed squared of the fluctuations
$c_s^2$ becomes negative. This
occurs independent of the above catastrophe, and leads to instabilities
of the perturbations at small scales. Even a very short period of $c_s^2<0$ would
 result in a disaster, since the growth rate of a mode is proportional to the wave 
number $k$, and all modes with wavelengths smaller than $\sqrt{-c_s^2}\Delta t$ 
(where $\Delta t$ is the duration of $c_s^2<0$), at least up to the Planck scale, 
would grow exponentially as $\exp(k\sqrt{-c_s^2}\Delta t)$ during that period.
We have to resolve this issue in order to obtain a consistent inflationary scenario.

These instabilities can be avoided by tuning the self-coupling coefficient $\lambda$ to a small value. 
The singularity of the equation of motion can be avoided in case
\begin{equation}
 \lambda<7.2\times10^{-9},
\end{equation}
while $c_s^2>0$ is maintained if\footnote{We did not reproduce the
result from \cite{Ohashi:2012wf}, which claims the upper bound of the Higgs 
quartic coupling $\lambda$ as $\lambda<2.7 \times 10^{-8}$.}
\begin{equation}
 \lambda < 2.9 \times 10^{-9}.
\end{equation}
This corresponds to making the value of $M$ larger, which in turn
diminishes the effects of the Galileon term after inflation. But we are
now considering the SM Higgs field, which has $\lambda$
determined by collider experiments, much larger than the above
value. Thus, we consider ameliorating these issues by extending the
structure of the Lagrangian in the next section.

%%%%%%%%%%%%%%%%%%%%%%%%%%%%%%%%%%%%%%%%%%%%%%%%
%%%%%%%%%%%%%%%%%%%%%%%%%%%%%%%%%%%%%%%%%%%%%%%%

\section{Extending the Higgs G-inflation model}
\label{sec3}

In this section, we extend the Lagrangian of the Higgs G-inflation
model, in order to avoid the instabilities stated in the previous
section. 
Furthermore, we aim to construct
a model with predictions preferred by the Planck satellite results.

For the detailed calculations, we refer to the formulation of
Generalized G-inflation \cite{Kobayashi:2011nu}.

\subsection{Adding a higher-order kinetic term}

The situation can be alleviated by adding a higher-order kinetic term to
the Higgs field Lagrangian.  We illustrate it first by adding a term of
the form
\begin{equation} 
\frac{1}{2\widetilde{M}^4}X^2,
\end{equation}
to the Higgs field Lagrangian. Here $\widetilde{M}$ represents the
scale of new physics associated with this term. The resultant action
becomes,
\begin{equation}
S= \int d^4 x \sqrt{-g}\left[ \frac{M_{P}^2}{2}R + X +
			\frac{1}{2\widetilde{M}^4}X^2 -
			\frac{1}{4}\lambda \varphi^4 + \frac{\varphi X}{M^4}
			\Box \varphi \right]. 
\end{equation}
By adding this term, from Eq.~(\ref{back13}) with $\ell=2$, $m=1$ and $n=0$,
the full equation of motion for the homogeneous part becomes
\begin{align}
 \left(1+\frac{3\dot{\varphi}^2}{2\widetilde{M}^4}-6H\frac{\varphi\dot{\varphi}}{M^4}+2\frac{\dot{\varphi}^2}{M^4} 
 + \frac{3\varphi^2\dot{\varphi}^4}{2M_P^2M^8}\right)\ddot{\varphi} +3\left(1+\frac{{\dot \varphi}^2}{2 {\widetilde M}^4}\right)H\dot{\varphi} +\lambda \varphi^3& \notag\\
-\left(9H^2 - \frac{3\dot{\varphi}^4}{8M_P^2\widetilde{M}^4}-\frac{3\dot{\varphi}^2}{2M_P^2}-\frac{3\dot{\varphi}^4}{2M_P^2M^4}+\frac{3\lambda\varphi^4}{4M_P^2}\right)& \frac{\varphi \dot{\varphi}^2}{2M^4} =0.
\end{align}
We see that a positive term $\frac{3}{2\widetilde{M}^4}\dot{\varphi}^2$,
which comes from the higher-order kinetic term, is added to the
coefficient of $\ddot{\varphi}$. One expects that this term makes the
coefficient positive throughout the reheating phase. The same type of term comes in when we add an arbitrary power of $X$ to the Lagrangian, and thus those terms can also be used.

The sound speed squared of the fluctuations is also modified as
\begin{equation}
c_s^2 = 
\frac{1 + \frac{1}{2\widetilde{M}^4}\dot{\varphi}^2 
      - (4H\dot{\varphi}+2\ddot{\varphi})\frac{\varphi}{M^4}
      - \frac{\varphi^2 \dot{\varphi}^4}{2M_{Pl}^2M^8} }
     {1 + \frac{3}{2\widetilde{M}^4}\dot{\varphi}^2
      - 6H\frac{\dot{\varphi}\varphi}{M^4}
      + 2 \frac{\dot{\varphi}^2}{M^4}
      + \frac{3\varphi^2 \dot{\varphi}^4}{2M_{Pl}^2M^8} }.
\label{cs2}
\end{equation}
One finds a positive contribution both in the numerator and in the denominator,
which is expected to help avoid negative values of $c_s^2$.

We carried out numerical calculations of the background evolution, using
the equation of motion of the scalar field and the gravitational evolution equation 
[the explicit form shown in Eq.~(\ref{back12}) with $\ell=2$, $m=1$, $n=0$]. The parameters $M$ and $\widetilde{M}$ were determined
by requiring the amplitude of the curvature fluctuations
to be $2.2\times 10^{-9}$ at
$k=0.05\mathrm{Mpc}^{-1}$ \cite{Ade:2013zuv}, which we have tentatively identified with the comoving Hubble scale 60 e-folds
before the end of inflation. The case in which this scale corresponds to different e-folds will be shown later. The perturbation values were determined
using the general formulas of Generalized G-inflation \cite{Kobayashi:2011nu}.

\begin{figure}[!h]
\centering
\includegraphics[width=11cm,clip]{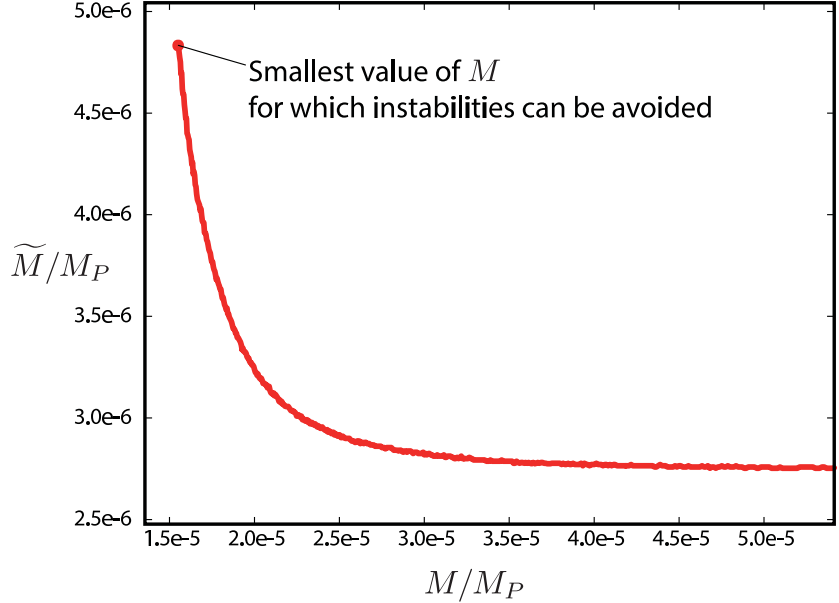}
\caption{The relation between $M$ and ${\widetilde M}$ that reproduces the correct amplitude of the power spectrum of the primordial scalar perturbation. For $M<1.6 \times 10^{-5} M_{P}$, the Galileon term induces instability at small scales during the oscillating stage, and the present universe will not be realized. }
\label{fig:mtildem}
\end{figure}

We have two parameters $M$ and $\widetilde{M}$ to tune the power
spectrum amplitude, so we obtain a one parameter family of possible
values for the parameters. 
Figure~\ref{fig:mtildem} shows the relation between $M$ and ${\widetilde M}$ that reproduces the correct amplitude 
of the power spectrum of the primordial scalar perturbation.
Instabilities are avoided when
\begin{equation}
 M>1.6\times 10^{-5}M_P.
\end{equation}
$\widetilde{M}$ is determined by the value of $M$, which for the smallest value of $M$ is
\begin{equation}\widetilde{M} \simeq 4.8 \times 10^{-6} M_P.\end{equation}
For smaller values of $M$, the effect of the Galileon term
after inflation becomes too strong compared to the kinetic term, leading
to instabilities. Slow-roll inflation can be realized even when
$M \rightarrow \infty$, which corresponds to
\begin{equation}
 \widetilde{M} = 2.7 \times 10^{-6}M_P. \label{kindom}
\end{equation}

For the smallest values of $M$, or $(M,{\widetilde M})=(1.6 \times 10^{-5} M_P, 4.8 \times 10^{-6} M_P)$, 
the spectral index becomes
\begin{equation}
 n_s \simeq 0.964,
\end{equation}
and the tensor-to-scalar ratio is
\begin{equation}
 r \simeq 0.155.
\end{equation}
Raising the value of $M$, both the tensor-to-scalar ratio and the
scalar spectral index become smaller. For $M\rightarrow \infty$, we find
\begin{equation}
 n_s \simeq 0.958,
\end{equation}
\begin{equation}
 r \simeq 0.116.
\end{equation}

Although the Lagrangian leads to a consistent cosmology, there is no reason for 
the specific $X^2$ term to dominate over the other higher-order kinetic terms 
possibly present. In the next section, we generalize the above model and explore its predictions.

Before we proceed, we would like to comment on another possible term that could be added, which is of the form
\begin{equation}
\varphi^2 X.
\end{equation}
This term also gives a positive contribution to both the coefficient of $\ddot{\varphi}$ in the equation of motion, and the sound speed squared. In the case in which this term dominates, this model becomes identical to running kinetic inflation \cite{Nakayama:2010kt} (see \cite{Cai:2012va} for another possible role of this type of term in the bouncing scenario).
Here, we do not pursue this possibility and concentrate on the higher-order kinetic term.

%%%%%%%%%%%%%%%%%%%%%%%%%%%%%%

\subsection{Generalized Lagrangian and its predictions}
Here we further analyze the predictions the model entails by using a generalized version of the above model.
The full action we consider is of the form
\begin{equation}
S= \int d^4 x \sqrt{-g}\left[ \frac{M_{P}^2}{2}R +k(X)
			- V(\varphi)
			+g(\varphi) h(X) \Box
			\varphi \right].
\end{equation}
The formulas for the background equations of motion and the perturbations are presented in Appendix A.

Let us consider the inflationary stage in which
the energy density is dominated by the potential term, and
in which the slow-roll parameters defined as
\begin{align}
\epsilon:=-\frac{\dot H}{H^2},
\quad
\eta:=-\frac{\ddot\varphi}{H\dot\varphi},
\quad
\alpha:=\frac{\dot g}{H g},
\end{align}
satisfy
\begin{align}
\epsilon,\;|\eta|,\;|\alpha|\ll 1.
\end{align}
The Friedmann equation~(\ref{back1})
reduces to
\begin{align}
3M_P^2H^2\simeq V,
\end{align}
while Eq.~(\ref{back2})
implies that
\begin{align}
k, \; Xk_X,\;  Hgh\dot\varphi \lesssim  M_P^2H^2\times {\cal O}(\epsilon).
\end{align}
Eq.~(\ref{back31}) in the slow-roll approximation is given by
\begin{align}
3k_X H\dot\varphi +V'
-18H^2ghm
\simeq 0,
\end{align}
where we have used
\begin{align}
\frac{2X g''}{H^2g}=\left(-\epsilon + \eta  + \alpha+\frac{\dot\alpha}{H\alpha}\right)\alpha\ll 1.
\end{align}
and introduced $m$ defined as
\begin{align}
m(X):=\frac{Xh_X}{h}.
\end{align}
Similarly, from Eqs.~(\ref{pertFs}) and (\ref{pertGs}) we have
\begin{align}
{\cal F}_S&\simeq \frac{1}{M_P^2 H^2}\left(
Xk_X-4Hgh\dot\varphi m
\right),
\\
{\cal G}_S&\simeq\frac{1}{M_P^2H^2}\left[
Xk_X+2X^2k_{XX}-6 Hgh\dot\varphi
\left(m^2+Xm_X\right)
\right].
\end{align}

We now consider two extreme cases analytically, one in which the kinetic term dominates the
dynamics of the Higgs field, and the other in which the Galileon term
dominates the dynamics of the Higgs field.
For $Xk_X, X^2k_{XX}\gg Hgh\dot\varphi$, which is the case where the
kinetic term dominates, the known result of k-inflation
\cite{ArmendarizPicon:1999rj, Garriga:1999vw} is reproduced,
\begin{align}
{\cal F}_S\simeq \epsilon,\quad
c_s^2\simeq \frac{k_X}{k_X+2Xk_{XX}},
\end{align}
where the background equation was used to derive the first equation.
The power spectrum of $\zeta$ is given by
\begin{align}
{\cal P}_\zeta\simeq\frac{1}{8\pi^2 c_s\epsilon}\left(\frac{H}{M_P}\right)^2,
\quad
n_s-1\simeq -2\epsilon -\frac{\dot {c_s}}{Hc_s}-\frac{\dot\epsilon}{H\epsilon} \label{ns_kin},
\end{align}
while the tensor-to-scalar ratio is $r\simeq 16c_s\epsilon$.  Note in
passing that $\dot \epsilon/H\epsilon\simeq 2\epsilon-\eta -
\eta/c_s^2$.

In the opposite limit, $Xk_X, X^2k_{XX}\ll Hgh\dot\varphi$, we have
\begin{align}
{\cal F}_S\simeq \frac{4}{3}\epsilon,\quad
c_s^2\simeq \frac{2}{3(m+Xm_X/m)}. \label{cs_m}
\end{align}
We are in particular interested in the model with $h\propto X^m$ where $m={\mathrm{const.}}$
In this case, $c_s^2\simeq 2/(3m)$ and
thus
\begin{align}
{\cal P}_\zeta\simeq\frac{1}{8\pi^2 \epsilon}\frac{3\sqrt{6m}}{8}
\left(\frac{H}{M_P}\right)^2,
\quad
n_s-1\simeq -2\epsilon -\frac{\dot\epsilon}{H\epsilon},
\quad
r\simeq \frac{64}{9}\sqrt{\frac{6}{m}}\epsilon.
\end{align}
Since $\dot \epsilon/H\epsilon \simeq \epsilon+\alpha-(2m+1)\eta$, we find
\begin{align}
n_s-1\simeq -3\epsilon-\alpha+(2m+1)\eta. \label{ns_gal}
\end{align}
For $m=1$ the result of the original Higgs G-inflation
\cite{Kamada:2010qe} is reproduced.\\

More concretely, we concentrate on the model with $g\propto
\varphi^{2n+1}$ and $h\propto X^m$. We take the following ansatz during
inflation:
\begin{align}
\epsilon=\frac{b}{N+b},
\quad
\eta=\frac{c}{N+b},
\quad
\alpha=\frac{d}{N+b}, \label{slow_def}
\end{align}
with $b,c$, and $d$ being constant, 
which will be justified later. Here, $N$ is the number of e-folds
defined by $dN:=-Hdt$ and inflation ends at $N=0$.  Then, we find
\begin{align}
H\sim (N+b)^b,\quad
X\sim (N+b)^{2c},
\quad
\varphi\sim (N+b)^{c-b+1},
\quad
g(\varphi)\sim (N+b)^{-d},
\end{align}
so that
\begin{align}
b=2(c-b+1),\quad
(2n+1)(c-b+1)=-d
\quad\Rightarrow\quad
c=\frac{3}{2}b-1,\quad
d=-\frac{2n+1}{2}b.
\end{align}

If $Xk_X \gg Hgh\dot\varphi$ and a single term $X^\ell$ dominates over the other terms in $k$,
then, we find from the slow-roll equation of motion
\begin{align}
3(c-b+1)=(2\ell -1)c+b.
\end{align}
We therefore arrive at
\begin{align}
b=\frac{2\ell-1}{3\ell-2}, \ c=\frac{1}{2(3\ell -2)}, \ d=-\frac{(2n+1)(2\ell-1)}{2(3\ell-2)}.
\label{eq:bcd}
\end{align}
Thus, from Eqs.~(\ref{ns_kin}) and (\ref{slow_def}), we obtain
\begin{equation}n_s-1 = -\frac{7\ell-4}{(3\ell-2)N+2\ell-1},\end{equation}
where we used
\begin{equation}c_s^2 \simeq \frac{1}{2\ell-1}\simeq \mathrm{const.} \label{cs_l}\end{equation}
The tensor-to-scalar ratio is
\begin{equation}r=16c_s\epsilon \simeq \frac{16\sqrt{2\ell-1}}{(3\ell -2)N +2\ell -1}.\end{equation}

If $Xk_X \ll Hgh\dot\varphi$, again, we find from the slow-roll equation of motion
\begin{align}
3(c-b+1)=2b+(2n+1)(c-b+1)+2mc,
\end{align}
leading to
\begin{align}
b=\frac{2m}{3m+n+1}, \ \ c=-\frac{n+1}{3m+n+1}, \ \
 d=-\frac{m(2n+1)}{3m+n+1}.
\label{eq:bcd2}
\end{align} 
Using Eqs.~(\ref{ns_gal}) and
(\ref{slow_def}), we obtain
\begin{equation}n_s-1 \simeq -\frac{7m+n+1}{(3m+n+1)N+2m},\end{equation}
and
\begin{equation}r\simeq\frac{64}{9}\sqrt{\frac{6}{m}}\epsilon \simeq \frac{128\sqrt{6m}}{9\left[(3m+n+1)N + 2m\right]}.\label{analyticr}\end{equation}
Note that Eqs.~(\ref{eq:bcd}) and (\ref{eq:bcd2}) show that 
the constants $b, c,$ and $d$ do not depend on $N$ and hence justify the
assumption (\ref{slow_def}).
The analytical predictions for these models are presented in Fig.~\ref{fig:prediction}.
Calculations using explicit Lagrangians are presented in Appendix B.

The model parameters $\ell, m,$ and $n$ intoduced here are in general arbitrary, but since the sound speed during inflation becomes smaller the larger these parameters are, there is an upper bound on the parameters due to the non-Gaussianities produced.
From the Planck constraints on non-Gaussianity \cite{Ade:2013ydc}, we have a lower bound on the sound speed during inflation,
\begin{equation}
c_s \gtsim 0.02.
\end{equation}
Comparing this with (\ref{cs_m}) and (\ref{cs_l}), we obtain an upper bound on the model parameters $\ell$ and $m$:
\begin{equation}
\ell, \ m \ltsim \mathcal{O}(10^3).
\end{equation}

\begin{figure}[ht]
\centering
\includegraphics[width=13cm,clip]{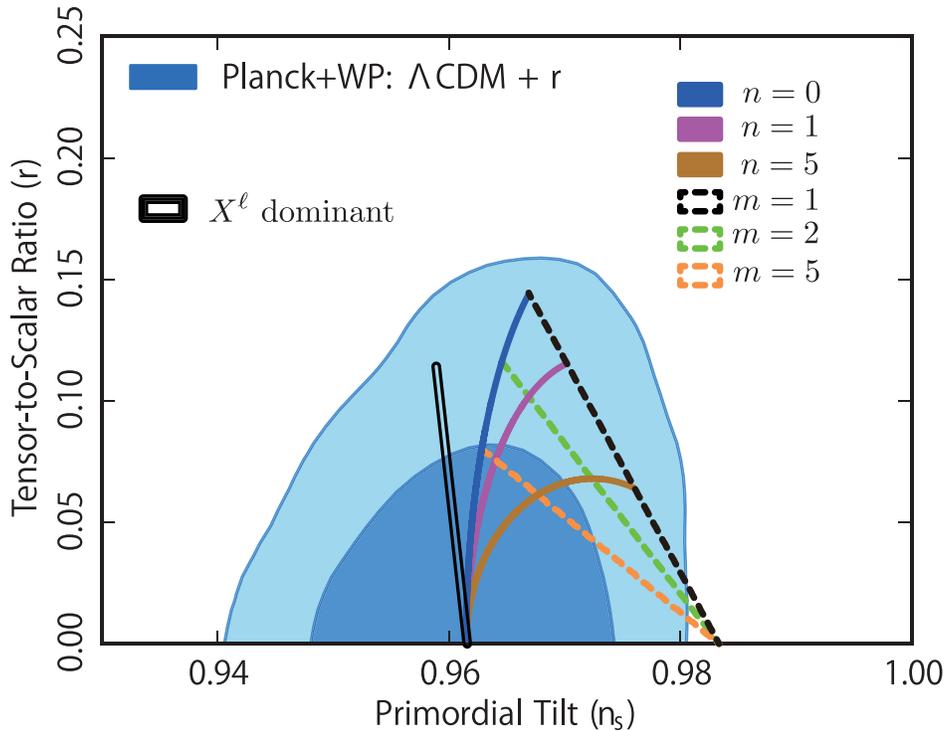}
\caption{Analytical predictions of the Galileon term dominating models and the higher-order kinetic term dominating models, compared with the Planck results \cite{Ade:2013uln}. The double line shows the case where the higher-order kinetic term is dominant, with varying $\ell$, while the other lines show the predictions of the Galileon term dominating case, fixing $n$ and varying $m$ or vice versa. }
\label{fig:prediction}
\end{figure}

%%%%%%%%%%%%%%%%%%%%%%%%%%%%%

\subsection{Numerical Analysis}

We now resort to numerical calculations to see which term dominates during inflation, while maintaining stability. For the Higgs field action, we used
\begin{equation}S = \int d^4x\sqrt{-g}\left[\frac{M_P^2}{2} +X + \frac{X^\ell}{\ell \widetilde{M}^{4(\ell-1)}} - \frac{1}{4}\lambda \varphi^4 + \frac{\varphi^{2n+1}X^m}{M^{2n+4m}}\Box \varphi\right].\end{equation}

We obtained the sets of $M$ and $\widetilde{M}$ with $\mathcal{P}_\zeta=2.2\times 10^{-9}$ at $N=60$, down to the smallest value of $M$ possible without encountering instabilities.
\begin{figure}[!h]
\centering
\includegraphics[width=12cm,clip]{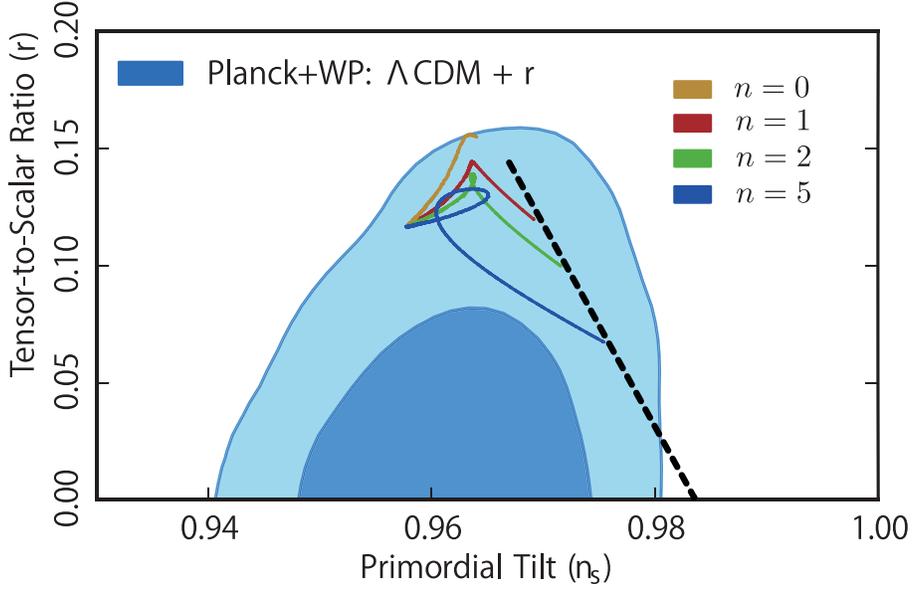}
\caption{Numerical results for $\ell=2$ and $m=1$ with several values of $n$. The continuous lines show the predictions for different combinations of $M$ and $\widetilde{M}$ that lead to $\mathcal{P}_\zeta=2.2\times10^{-9}$ at $N=60$ while avoiding instabilities. The black line is the analytical prediction of the Galileon term dominant case.}
\label{fig:prediction1}
\end{figure}
\begin{figure}[!h]
\centering
\includegraphics[width=12cm,clip]{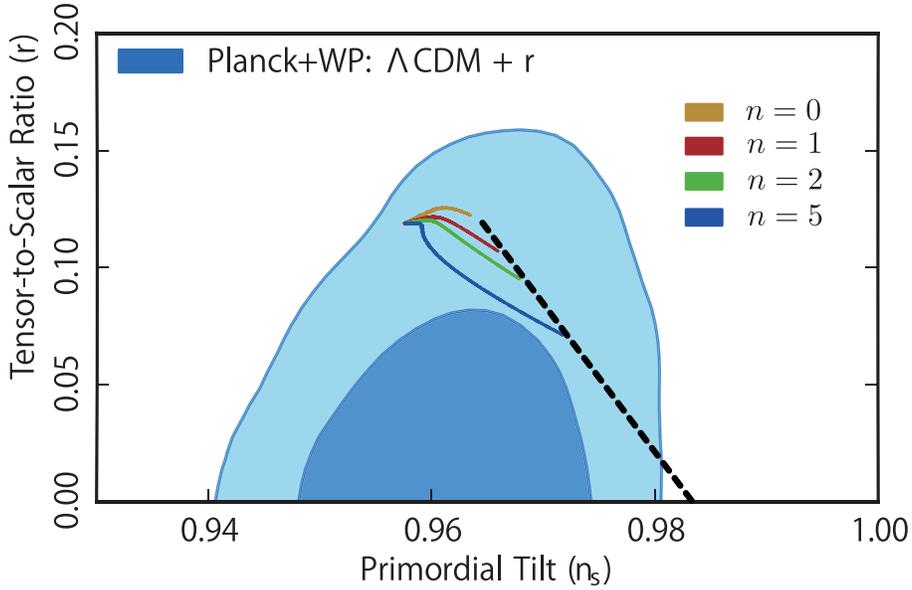}
\caption{Results for $\ell=2, m=2$ with several values of $n$, varying the value of $M$.}
\label{fig:prediction2}
\end{figure}
\begin{figure}[!h]
\centering
\includegraphics[width=12cm,clip]{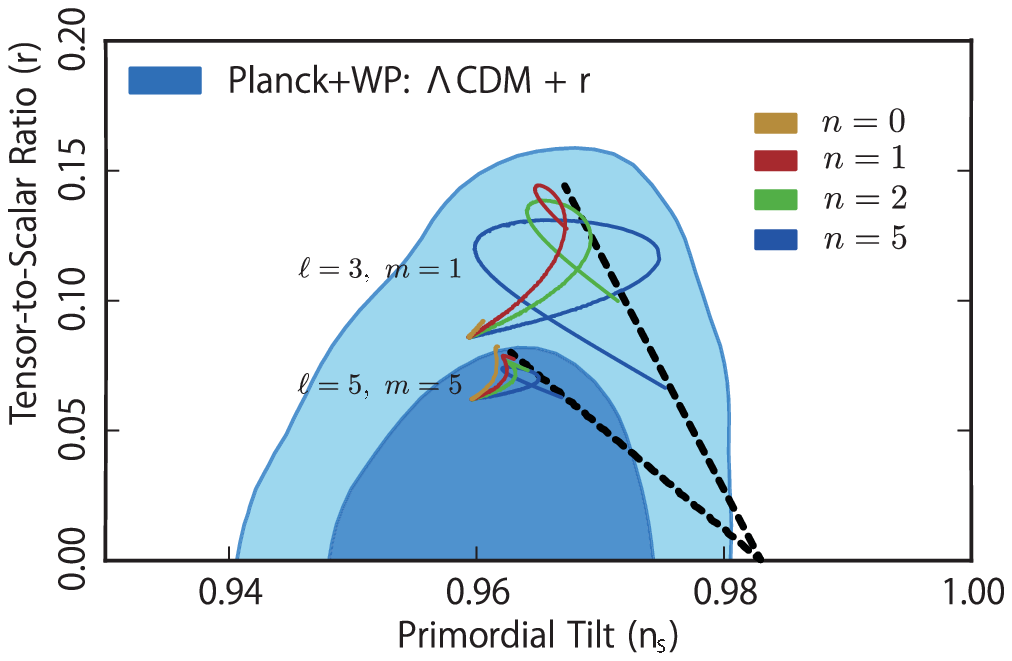}
\includegraphics[width=12cm, clip]{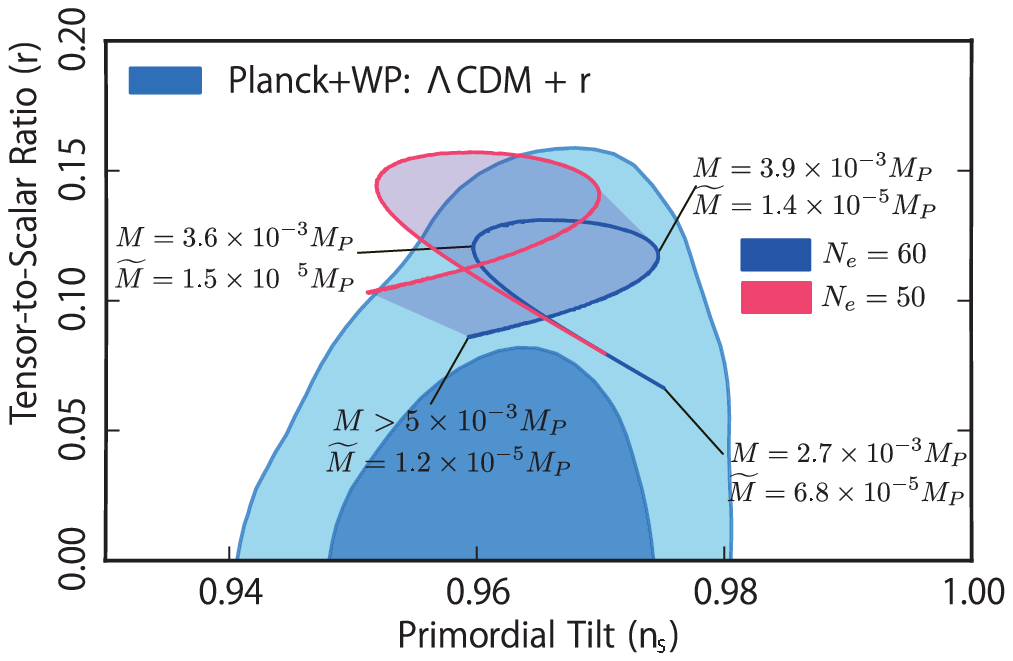}
\caption{Results for $\ell=3,\ m=1$, and $\ell=5,\ m=5$ for $N=60$
 (top), and $\ell=3,\ m=1,\ n=5$ with different numbers of e-folds
 (bottom). The value of $M$ for each point is shown in the lower
 figure. }
\label{fig:prediction3}
\end{figure}
For $\ell=2$ and $m=1$, the results are shown in 
Fig.~\ref{fig:prediction1}. The focal point at $(n_s, r) = (0.958, 0.116)$
corresponds to the case where the dynamics is dominated by the
higher-order kinetic term with no Galileon effect. This corresponds to
the upper end of the double line in Fig.~\ref{fig:prediction}, but the
actual numerical values are slightly deviated due to the crudeness of
the analytic estimate used to draw Fig.~\ref{fig:prediction}. The fact
that the $n=0$ line corresponding to the original Higgs G-inflation is
discontinued before reaching the black solid line in 
Fig.~\ref{fig:prediction1} indicates that the Galileon dominant case beyond
the end point is not viable due to instabilities after inflation. We see
that for large values of $n$, the Galileon term can dominate the Higgs
field dynamics during inflation, recovering the analytical prediction of
the Galileon dominant case, while for $n=0$, the higher-order kinetic
term comes into effect whenever instabilities are avoided.

We also show several numerical results for other parameters in Figs.~\ref{fig:prediction2} 
and \ref{fig:prediction3}.  These models
typically predict a spectral index consistent with the Planc]k results,
and large values for the tensor-to-scalar ratio.
In conclusion, our numerical calculation shows that 
introducing a higher-order kinetic term as well as a higher-order Galileon term 
helps avoid the unwanted instability during the reheating stage of Higgs G-inflation,
while its predictions come to parameter regions that can be
tested by the CMB observations in the near future.

\section{Conclusion}

In this paper, we extended the Lagrangian of the Higgs G-inflation model
by adding a higher-order kinetic term and modifying the form of the
Galileon term.  By adding the higher-order kinetic term, both the
coefficient of $\ddot{\varphi}$ in the equation of motion and the sound
speed squared receive a positive contribution, thereby avoiding the
instabilities reported in Ref.~\cite{Ohashi:2012wf}. Modifying the Galileon
term enables the term to dominate over the higher-order kinetic term in
the equation of motion during inflation, leading to a new class of
models.

As shown in Figs.~\ref{fig:prediction1} - \ref{fig:prediction3}, our
model matches the observed range of the spectral index quite well, while
the tensor-to-scalar ratio takes values between the 1$\sigma$ and 2$\sigma$
bounds for the parameter range we probed, unlike the original Higgs
inflation model. This is good news for those working in B-mode
polarization experiments.

To suppress the amplitude of $r$, one should take either large
$m$ or large $n$.  In the limit of large $n$, the spectral index
converges to $n_s=1-1/N\simeq 0.983$ for $N=60$, which is outside the
2$\sigma$ range. On the other hand, in the large $m$ limit, $n_s$
approaches $ n_s = 1- \frac{7}{3(N+2/3)}\simeq 0.962$ for $N=60$, which
is in the preferred range. Hence, to suppress $r$ in this
model, we can introduce higher-order interactions with large
exponents. Our model thus accommodates a large region of the $n_s$ - $r$
plane, most of which is consistent with current observations.

\section*{Acknowledgements}

This work was supported in part by 
% JSPS postdoctoral fellowship for research abroad (K.K), 
the JSPS Grant-in-Aid for Young
Scientists (B) No.~24740161 (T.Ko.), the Grant-in-Aid for Scientific
Research on Innovative Areas No.~24111706 (M.Y.) and No.~21111006 (J.Y.), and the Grant-in-Aid
for Scientific Research No.~25287054 (M.Y.) and No.~23340058 (J.Y.), and the Program for Leading Graduate
Schools, MEXT, Japan (T.Ku.).
\\

%%%%%%%%%%%%%%%%%%%%%%%%%%%%%%%%%%%%%%%%
%%%%%%%%%%%%%%%%%%%%%%%%%%%%%%%%%%%%%%%%
%%%%%%%%%%%%%%%%%%%%%%%%%%%%%%%%%%%%%%%%
%%%%%%%%%%%%%%%%%%%%%%%%%%%%%%%%%%%%%%%%

\appendix
{\Large {\bf Appendixes}}

\section{General formulas}
In this appendix, we give the general formulas for the equations of
motion and primordial curvature perturbations.

The general cosmological background and perturbation equations for the
Lagrangian
\begin{align}
{\cal L}=\frac{M_P^2}{2}R+K(\varphi, X)-G(\varphi, X)\Box\varphi
\end{align}
are found in Refs.~\cite{Deffayet:2010qz, Kobayashi:2010cm, Kobayashi:2011nu}.
The background equation of motion for the scalar field $\varphi$ is given by
\begin{align}
&\lefteqn{\left(K_X+2XK_{XX}+6H\dot{\varphi}G_X+6H\dot{\varphi}XG_{XX}-2G_{\varphi}
  -2XG_{X\varphi} \right)\ddot{\varphi}} \notag \\
 &+3H\dot{\varphi}K_X - K_{\varphi} + 2(9H^2+3\dot{H})XG_X - 6 H
 \dot{\varphi}G_\varphi -2XG_{\varphi\varphi}+6H\dot{\varphi} XG_{X\varphi} + 2X K_{\varphi X}=0,
\label{eqn:scalar}
\end{align}
where the subscripts on $K$ and $G$ denote the derivative with respect to
those variables. The gravitational field equations are given by
\begin{align}
3M_{P}^2H^2 &= 2XK_X-K+6H\dot{\varphi}XG_{X}-2XG_{\varphi}, 
\label{eqn:grav1} \\
M_{P}^2\dot{H} &=
 -XK_X-3H\dot{\varphi}XG_{X}+2XG_{\varphi}+\ddot{\varphi}XG_X.
\label{eqn:grav2}
\end{align}

In the main text we focus on inflation models for which
the functions $K$ and $G$ are of the form
\begin{align}
K=k(X)-V(\varphi),\quad G=-g(\varphi)h(X).\label{kgh}
\end{align}

To study the potential-driven inflation in the above theory,
it is convenient to introduce
\begin{align}
\epsilon:=-\frac{\dot H}{H^2},
\quad
\eta:=-\frac{\ddot\varphi}{H\dot\varphi},
\quad
\alpha:=\frac{\dot g}{H g}.\label{slowpar}
\end{align}
The background gravitational field equations are now given by
\begin{align}
3M_{P}^2 H^2&=2Xk_X-k+V-Hgh\dot\varphi\left(6m-\alpha\right),\label{back1}
\\
M_{P}^2\dot H&=
-Xk_X+Hgh\dot\varphi\left(3m+m\eta -\alpha\right),\label{back2}
\end{align}
where we also introduced
\begin{align}
m(X):=\frac{Xh_X}{h}.
\end{align}
The background scalar-field equation is
\begin{align}
0=&H\dot{\varphi}\bigg{[}k_X(3-\eta) +H\dot{\varphi}gh_X\left(-9-3\alpha +6\eta +3\epsilon - \alpha\eta\right) \notag\\
&+2Hg\frac{h}{\dot{\varphi}}(3-\eta)\alpha-2Xk_{XX}\eta +6H\dot{\varphi}Xgh_{XX}\eta \bigg{]}
+V'\left(1+\frac{2Xg''h}{V'}\right) \label{back31}\\
=&A\ddot\varphi+3k_X H\dot\varphi +V'\notag\\
&+H^2gh\left[6m\left(-3+\frac{Xk_X}{M_P^2H^2}- \frac{3mgh\dot{\varphi}}{M_P^2H}\right)+6\alpha\left(\frac{mgh\dot{\varphi}}{M_P^2H}-m+1\right)+\frac{2X g''}{H^2g}\right],\label{back3}
\end{align}
where
\begin{align}
A=k_X+2Xk_{XX}-6\frac{Hgh\dot\varphi}{X}\left(m^2+Xm_X\right)+\frac{Hgh\dot\varphi}{X}(m+1)\alpha + \frac{6m^2g^2h^2}{M_P^2}.
\end{align}
Here, we have eliminated $\dot{H}$ in Eq.~(\ref{back3}) using Eq.~(\ref{back2}).  If $A(t_*)=0$
at some $t=t_*$, we cannot solve the time evolution of $\varphi$ for
$t>t_*$.  Therefore, we require that $A$ never crosses the zero.  Note
here that we have not employed any slow-roll approximations; the above
equations can be used in the reheating stage as well as during
inflation.

The second-order action $S_2$ for the primordial curvature perturbation
$\zeta$ is given by \cite{Kobayashi:2011nu}
\begin{equation}
  S_2 = M_P^2 \int d^4 x \,a^3\left[\mathcal{G}_S \dot{\zeta}^2
	-\frac{\mathcal{F}_S}{a^2}(\vec{\nabla}\zeta)^2\right],
\end{equation}
where
\begin{align}
\mathcal{F}_S &= \frac{M_P^2X}{\Theta^2}\bigg{[}K_X - 2 G_\varphi + 4
 H \dot{\varphi} G_X + 2 \ddot{\varphi} G_X +2X \ddot{\varphi} G_{XX} + 2X G_{X\varphi} - \frac{2}{M_P^2}X^2 G_X^2\bigg{]}\\
&=\frac{M_P^2X}{\Theta^2}\left\{
k_X+\frac{Hgh\dot\varphi}{X}\left[-4m-(m-1)\alpha+2(m^2+Xm_X)\eta\right]
-\frac{2}{M_P^2}g^2h^2m^2
\right\},
\\
{\cal G}_S&= \frac{M_P^2X}{\Theta^2}\bigg{[}K_X+2XK_{XX} - 2
 G_\varphi + 6 H \dot{\varphi} G_X +6 H \dot{\varphi}X G_{XX} - 2X G_{X\varphi} + \frac{6}{M_P^2}X^2 G_X^2\bigg{]}\\
&=\frac{M_P^2X}{\Theta^2}\left\{
k_X+2Xk_{XX}+\frac{Hgh\dot\varphi}{X}\left[
-6\left(m^2+Xm_X\right)+(m+1)\alpha\right]
+\frac{6}{M_P^2}g^2h^2m^2
\right\},
\end{align}
and
\begin{align}
\Theta = M_P^2H\left(1+m\frac{Hgh\dot\varphi}{M_P^2H^2}\right).
\end{align}
The sound speed is given by $c_s^2={\cal F}_S/{\cal G}_S$.
The curvature perturbation shows a stable evolution provided that ${\cal F}_S>0$
and ${\cal G}_S>0$.
Note again that no slow-roll approximations are made in deriving the above expressions,
and hence the same conditions ${\cal F}_S>0$ and ${\cal G}_S>0$ can be used to
judge the stability during the reheating stage.

%----------------------------------

%%%%%%%%%%%%%%%%%%%%%%%%%%%%%%%%%
%%%%%%%%%%%%%%%%%%%%%%%%%%%%%%%%%

\section{Analytical calculations}
\label{sec4}

In this section, we give analytic formulas using an explicit action of the form
\begin{equation}S = \int d^4x\sqrt{-g}\left[\frac{M_P^2}{2}R + X + \frac{X^\ell}{\ell \widetilde{M}^{4(\ell-1)}} - \frac{1}{4}\lambda \varphi^4 + \frac{\varphi^{2n+1}X^m}{M^{2n+4m}}\Box \varphi\right]\end{equation}
for the two extreme cases; one
in which the higher-order kinetic term dominates the equation of
motion, and the other in which the Galileon term dominates the equation
of motion. This action corresponds to taking 
\begin{equation}g(\varphi) = \varphi\left(\frac{\varphi^2}{M^2}\right)^n,\ \ h(X) = \left(\frac{X}{M^4}\right)^m,\end{equation}
and the term proportional to $X^\ell$ is to be understood as the dominant higher-order kinetic term during inflation.

\subsection{Evolution equations}
From Eqs.~(\ref{back1}) - (\ref{back3}) the evolution equations for this action are 
\begin{align}
3M_{P}^2 H^2&=X + \frac{(2\ell-1)X^\ell}{\ell\widetilde{M}^{4(\ell-1)}}+\frac{1}{4}\lambda \varphi^4-\frac{H\varphi^{2n+1}\dot{\varphi}X^m}{M^{2n+4m}}\left(6m-\frac{(2n+1)\dot{\varphi}}{H\varphi}\right),\label{back11}
\\
M_{P}^2\dot{H}&=
-X -\frac{X^\ell}{\widetilde{M}^{4(\ell - 1)}}+\frac{H\varphi^{2n+1}\dot{\varphi}X^m}{M^{2n+4m}}\left(3m-\frac{m\ddot{\varphi}}{H\dot{\varphi}} -\frac{(2n+1)\dot{\varphi}}{H\varphi}\right),\label{back12}
\end{align}
\begin{align}
&\left(1+\frac{(2\ell-1)X^{\ell-1}}{\widetilde{M}^{4(\ell-1)}}-\frac{6m^2H\varphi^{2n+1}\dot{\varphi}X^{m-1}}{M^{2n+4m}}+\frac{2(2n+1)(m+1)\varphi^{2n}X^{m}}{M^{2n+4m}} +
\frac{6m^2\varphi^{4n+2}X^{2m}}{M_P^2M^{4n+8m}}\right)\ddot\varphi\notag\\
&+3H\dot\varphi\left(1+ \frac{X^{\ell-1}}{\widetilde{M}^{4(\ell-1)}}\right) +\lambda \varphi^3
+\frac{H^2\varphi^{2n+1}X^{m}}{M^{2n+4m}}\left[-9m+\frac{3mX}{M_P^2H^2}+ \frac{3mX^\ell}{\ell M_P^2\widetilde{M}^{4(\ell-1)}H^2}\right.\notag\\
&\left.+\frac{6m(2n+1)\varphi^{2n}X^{m+1}}{M_P^2M^{2n+4m}H^2}-\frac{6(2n+1)(m-1)\dot{\varphi}}{H\varphi}-\frac{3m\lambda\varphi^4}{4M_P^2H^2}+\frac{4n(2n+1)X}{H^2\varphi^2}\right]
=0,\label{back13}
\end{align}
The coefficients in $S_2$ become

\begin{equation}
\mathcal{F}_S=\frac{M_P^2X}{\Theta^2}\left[1 +
\frac{X^{\ell-1}}{\widetilde{M}^{4(\ell-1)}}
-\frac{H\varphi^{2n+1}\dot{\varphi}X^{m-1}}{M^{2n+4m}}\left(4m+\frac{(m-1)(2n+1)\dot{\varphi}}{H\varphi}+\frac{2m^2\ddot{\varphi}}{H\dot{\varphi}}\right)-\frac{2m^2\varphi^{4n+2}X^{2m}}{M_P^2M^{4n+8m}}\right],
\label{pertFs}
\end{equation}
\begin{equation}
\mathcal{G}_S=\frac{M_P^2X}{\Theta^2}\left[1 +
\frac{(2\ell-1)X^{\ell-1}}{\widetilde{M}^{4(\ell-1)}}-
\frac{H\varphi^{2n+1}\dot{\varphi}X^{m-1}}{M^{2n+4m}}\left(6m^2-\frac{(m+1)(2n+1)\dot{\varphi}}{H\varphi}\right)+\frac{6m^2\varphi^{4n+2}X^{2m}}{M_P^2M^{4n+8m}}\right],
\label{pertGs}
\end{equation}
with
\begin{equation}
\Theta=M_P^2H + \frac{m\varphi^{2n+1}\dot{\varphi}X^m}{M^{2n+4m}}.
\end{equation}
In the following, we analytically investigate the system introduced above for the two extreme cases.

\subsection{$X^\ell$ domination}

First, we investigate the case in which the higher-order kinetic term
dominates the inflation dynamics. This corresponds to the condition
\begin{equation}
k_X \gg |H\dot{\varphi}gh_X|,\ 1\label{galileonkinX}
\end{equation}
throughout inflation.

Assuming the slow-roll conditions,
\begin{equation}
|\epsilon|,\ |\eta|,\  |\alpha| \ll 1, \label{slow-rollX}
\end{equation}
Eqs.~(\ref{back11})-(\ref{back13}) can be approximated as
\begin{equation}
3M_P^2H^2 \simeq V = \frac14 \lambda \varphi^4, \label{friedmann}
\end{equation}
\begin{equation}
M_P^2\dot{H} \simeq
 -\frac{X^\ell}{\widetilde{M}^{4(\ell-1)}}, \label{friedmannX}
\end{equation}
\begin{equation}
3H\dot{\varphi}\frac{X^{\ell-1}}{\widetilde{M}^{4(\ell-1)}}+ \lambda \varphi^3 \simeq 0.
\end{equation}
Solving the last equation for $\dot{\varphi}$, we obtain
\begin{equation}
\dot{\varphi}  \simeq -\bigg{(}\sqrt{\frac{\lambda}{3}}2^\ell \widetilde{M}^{4(\ell-1)}M_P\varphi\bigg{)}^\frac{1}{2\ell-1}, \label{varphidotX}
\end{equation}
where use has been made of Eq.~(\ref{friedmann}) to eliminate $H$.

The number of e-folds $N$ can be calculated as
\begin{align}
N &= \int Hdt = \int \frac{H}{\dot{\varphi}} d\varphi \notag \\ 
&=-\int\sqrt{\frac{\lambda}{12}}\frac{\varphi^2}{M_{P}}\bigg{|}\sqrt{\frac{\lambda}{3}}2^\ell\widetilde{M}^{4(\ell-1)} {M_{P}}\varphi\bigg{|}^{-\frac{1}{2\ell-1}}d\varphi \notag \\
&= -\frac{2\ell-1}{3\ell-1}2^{-\frac{5\ell-2}{2\ell-1}}3^{-\frac{\ell-1}{2\ell-1}}\lambda^{\frac{\ell-1}{2\ell-1}}\widetilde{M}^{-\frac{{4(\ell-1)}}{2\ell-1}}M_{P}^{-\frac{2\ell}{2\ell-1}}\left[\varphi^{\frac{6\ell-4}{2\ell-1}}\right]^{\varphi_\mathrm{end}}_\varphi.
\end{align}
$\varphi_{\mathrm{end}}$ is given by the condition $\epsilon=1$, which reads
\begin{equation}
\varphi_{\mathrm{end}} =
2^{\frac{5\ell-2}{6\ell-4}}3^{\frac{\ell-1}{6\ell-4}}\lambda^{-\frac{\ell-1}{6\ell-4}}\widetilde{M}^{\frac{{4(\ell-1)}}{6\ell-4}}{M_{P}^{\frac{\ell}{3\ell-2}}}.
\end{equation}
This leads to
\begin{equation}
N =
 -\frac{2\ell-1}{3\ell-2}+\frac{2\ell-1}{3\ell-2}2^{-\frac{5\ell-2}{2\ell-1}}3^{-\frac{\ell-1}{2\ell-1}}\lambda^{\frac{\ell-1}{2\ell-1}}\widetilde{M}^{-\frac{{4(\ell-1)}}{2\ell-1}}{M_{P}^{-\frac{2\ell}{2\ell-1}}}\varphi^{\frac{6\ell-4}{2\ell-1}},
\end{equation}
and hence we express the value of $\varphi$ in terms of $N$:
\begin{equation}
\varphi = \left(\frac{3\ell-2}{2\ell-1}\right)^{\frac{2\ell-1}{6\ell-4}}2^{\frac{5\ell-2}{6\ell-4}}
 3^{\frac{\ell-1}{6\ell-4}}\lambda^{-\frac{\ell-1}{6\ell-4}}\widetilde{M}^\frac{{4(\ell-1)}}{6\ell-4}{M_{P}^{\frac{\ell}{3\ell-2}}}\left(N
+\frac{2\ell-1}{3\ell-2}\right)^{\frac{2\ell-1}{6\ell-4}}.
\end{equation}
Then, the slow-roll parameters can be expressed in terms of $N$ as
\begin{align}
 &\epsilon = -\frac{\dot{H}}{H^2}=-\frac{1}{2H}\frac{d}{dt}\ln H^2 \simeq
 -\frac{2\dot{\varphi}}{H\varphi}
 \simeq \frac{2\ell-1}{3\ell-2} \left(N+ \frac{2\ell-1}{3\ell-2}\right)^{-1}, \\
 &\eta = -\frac{\ddot{\varphi}}{H\dot{\varphi}} 
 =-\frac{1}{H}\frac{d}{dt}\ln|\dot{\varphi}| \simeq
 -\frac{\dot{\varphi}}{(2\ell-1)H\varphi}
 \simeq  \frac{1}{2(3\ell-2)} \left(N+ \frac{2\ell-1}{3\ell-2}\right)^{-1}.
\label{eq:X2slow}
\end{align} 
Thus, by an explicit calculation we have established the validity of
the ansatz~(\ref{slow_def}); $b=(2\ell-1)/(3\ell-2), c=1/2(3\ell-2)$.

The parameter $\widetilde{M}$ is determined by calculating the
curvature perturbations generated from this model. In the present case,
from Eqs.~(\ref{pertFs}) and (\ref{pertGs}), the coefficients in the
perturbation action $S_2$ read
\begin{align}
 \mathcal{F}_S &\simeq \frac{X^\ell}{H^2M_P^2\widetilde{M}^{4(\ell-1)}}, \nonumber \\   
 \mathcal{G}_S &\simeq \frac{(2\ell-1)X^\ell}{H^2M_P^2\widetilde{M}^{4(\ell-1)}}.
\end{align}
This leads to the sound speed
\begin{equation}
c_s^2 := \frac{\mathcal{F}_S}{\mathcal{G}_S} \simeq \frac{1}{2\ell-1}.
\end{equation}
The curvature
perturbations in the present case are expressed in terms of $N$ as
\begin{align}
  \mathcal{P}_\zeta &= \frac{1}{8\pi^2 c_s {\mathcal{F}_S}}
  \left(\frac{H}{M_P}\right)^2 \notag\\
 &= (3\ell-2)^{\frac{7\ell-4}{3\ell-2}}(2\ell-1)^{-\frac{11\ell-6}{6\ell-4}}2^{-\frac{5\ell-6}{3\ell-2}} 3^{-\frac{\ell}{3\ell-2}} \pi^{-2} \lambda^{\frac{\ell}{3\ell-2}}
  \left(\frac{\widetilde{M}}{M_P}\right)^{\frac{8(\ell-1)}{3\ell-2}} \left(N+\frac{2\ell-1}{3\ell-2}\right)^{\frac{7\ell-4}{3\ell-2}}.
\end{align}
Solving this for $\widetilde{M}$, we have
\begin{equation}
\widetilde{M}=\left(\mathcal{P}_\zeta \pi^2\right)^\frac{3\ell-2}{8(\ell-1)}
(3\ell-2)^{-\frac{7\ell-4}{8(\ell-1)}}(2\ell-1)^{\frac{11\ell-6}{16(\ell-1)}} 2^{\frac{5\ell-6}{8(\ell-1)}}3^{\frac{\ell}{8(\ell-1)}}\lambda^{-\frac{\ell}{8(\ell-1)}}\left(N+\frac{2\ell-1}{3\ell-2}\right)^{-\frac{7\ell-4}{8(\ell-1)}}M_P.
\end{equation}
For $\ell=2$, $N=60$ and
$\lambda=0.01$, we obtain
\begin{equation}\widetilde{M} \simeq 
 2.7\times 10^{-6}M_P\end{equation}
to explain the present universe with $\mathcal{P}_\zeta = 2.2 \times 10^{-9}$.

The spectral index and tensor-to-scalar ratio can also be calculated analytically.
The spectral index is calculated using the expression for the power spectrum:
\begin{equation}
n_s -1 = \frac{d\ln \mathcal{P}_\zeta }{d\ln k} 
\simeq \frac{1}{H}\frac{d}{dt}\ln \left(H^4 \dot{\varphi}^{-2\ell}\right)
=-4\epsilon + 2\ell\eta.
\end{equation}
For the tensor power spectrum, we obtain the same expression as in the usual canonical potential driven inflation:
\begin{equation}
\mathcal{P}_t = \frac{8}{M_P^2}\left(\frac{H}{2\pi}\right)^2.
\end{equation} 
Thus, the tensor-to-scalar ratio is calculated as
\begin{equation}
r:= \frac{\mathcal{P}_t}{\mathcal{P}_\zeta} =
 16 c_s \mathcal{F}_S \simeq
\frac{16\sqrt{2\ell-1}}{(3\ell-2)N+2\ell-1}.
\end{equation}
The tensor-to-scalar ratio becomes smaller for larger $\ell$.

%%%%%%%%%%%%%%%%%%%%%%%%%%%%%%%%%%%
%%%%%%%%%%%%%%%%%%%%%%%%%%%%%%%%%%%

\subsection{Galileon domination}

Next, we investigate the case in which the Galileon term dominates over the newly introduced higher-order kinetic term, namely
\begin{equation}
|H\dot{\varphi}gh_X| =\left|H\frac{m \varphi^{2n+1}
		      \dot{\varphi}^{2m-1}}{2^{m-1}M^{2n+4m}}\right|\gg
k_X\label{galileonkin}
\end{equation}
during inflation. This happens when we take large $n\gg1$. In this case,
the analytic formulas in Ref.~\cite{Kamada:2010qe} can be used with slight
modifications.

Assuming the slow-roll conditions,
\begin{equation}
|\epsilon|,\ |\eta|,\ |\alpha|  \ll 1, \label{slow-roll}
\end{equation}
Eqs.~(\ref{back11})-(\ref{back13}) can be approximated as
\begin{equation}
3M_P^2H^2 \simeq V = \frac14 \lambda \varphi^4, \label{friedmannG}
\end{equation}
\begin{equation}
M_P^2\dot{H} \simeq
 \frac{3mH\varphi^{2n+1}\dot{\varphi}^{2m+1}}{2^{m}M^{2n+4m}},
\end{equation}
\begin{equation}
\frac{9mH^2 \varphi^{2n+1} \dot{\varphi}^{2m}}{2^{m-1}M^{2n+4m}}+
 \lambda \varphi^3 \simeq 0.\label{sloweomG}
\end{equation}
Solving the last equation for $\dot{\varphi}$, we obtain
\begin{equation}
\dot{\varphi}  \simeq -\bigg{(}M_P \frac{2^{\frac{m+1}{2}}M^{n+2m}}{\sqrt{3m}\varphi^{n+1}}\bigg{)}^\frac{1}{m}, \label{varphidot}
\end{equation}
where we used Eq.~(\ref{friedmannG}) and took the negative sign for
$\dot{\varphi}$.

From the numerical calculations in the main text, we see that in order
to avoid instabilities after inflation, the higher-order kinetic term
has to dominate over the Galileon term by the end of inflation. This
situation makes analytical calculations substantially difficult, so here
we focus on the parameter regions where we can ignore the number of
e-folds of inflation during which the equation of motion is dominated by
the higher-order kinetic term. This is realized when we have a large
value of $n$, as seen from the numerical calculations. In this case, we
can approximately calculate the number of e-folds by assuming that the
Galileon term dominated until the end of inflation:
\begin{align}
N &= \int Hdt = \int \frac{H}{\dot{\varphi}} d\varphi \notag \\ 
&=-\int\sqrt{\frac{\lambda}{12}}\frac{\varphi^2}{M_P}\bigg{(}\frac{\sqrt{3m}\varphi^{n+1}}{M_P2^{\frac{m+1}{2}}M^{n+2m}}\bigg{)}^\frac{1}{m}d\varphi \notag \\
&= -\sqrt{\frac{\lambda}{12}}\frac{1}{M_P}\bigg{(}\frac{\sqrt{3m}}{M_P 2^{\frac{m+1}{2}}M^{n+2m}}\bigg{)}^\frac{1}{m}\frac{m}{3m + n+ 1}\bigg{[}\varphi^{\frac{3m+n+1}{m}}\bigg{]}^{\varphi_{\mathrm{end}}}_\varphi ,\label{efoldg}
\end{align}
where we used Eq.~(\ref{varphidot}). $\varphi_{\mathrm{end}}$ is
determined by the condition that
\begin{equation}
\epsilon = - \frac{\dot{H}}{H^2} \simeq \frac{9XH\dot{\varphi}G_X}{V}
=1,
\end{equation}
which gives
\begin{equation}
\varphi_{\mathrm{end}}\simeq \bigg{[}
 2^\frac{5m+1}{2m} 3^\frac{m-1}{2m} m^{-\frac{1}{2m}}
\lambda^{-\frac{1}{2}} M_P^\frac{m+1}{m}
 M^\frac{2m+n}{m} \bigg{]}^{\frac{m}{3m+n+1}}. \label{end}
\end{equation}
Inserting this value into Eq.~(\ref{efoldg}) leads to
\begin{equation}
 N = \frac{m}{3m+n+1}
     \left( 2^{-\frac{3m+1}{2m}} 3^{-\frac{m-1}{2m}} m^{\frac{1}{2m}}
            \lambda^{\frac{1}{2}} M_P^{-\frac{m+1}{m}} M^{-\frac{2m+n}{m}}
            \varphi^{\frac{3m+n+1}{m}} -2 \right),
\end{equation}
which yields the relation between the value of the Higgs field and
e-foldings $N$ before the end of inflation,
\begin{equation}
 \varphi = \left[ 2^{3m+1} 3^{m-1} m^{-1} \lambda^{-m} M_P^{2(m+1)}
     M^{2(2m+n)} \left( \frac{3m+n+1}{m}N+2 \right)^{2m}
     \right]^{\frac{1}{2(3m+n+1)}}. \label{varphiend}
\end{equation}
Then, the slow-roll parameters can be expressed in terms of $N$ as
\begin{align}
 &\epsilon = -\frac{\dot{H}}{H^2}=-\frac{1}{2H}\frac{d}{dt}\ln H^2 \simeq
 -\frac{2\dot{\varphi}}{H\varphi}
 \simeq \frac{2m}{N(3m+n+1)+2m}, \\
 &\eta = -\frac{\ddot{\varphi}}{H\dot{\varphi}} 
 =-\frac{1}{H}\frac{d}{dt}\ln|\dot{\varphi}| \simeq
 \frac{n+1}{mH}\frac{d}{dt}\ln \varphi
 \simeq -\frac{n+1}{N(3m+n+1)+2m}, \\
 &\alpha = \frac{{(2n+1)}\dot{\varphi}}{H\varphi} 
 \simeq {-}\frac{{(2n+1)}m}{N(3m+n+1)+2m}.
\label{eq:gslow}
\end{align} 

The parameter $M$ is determined by calculating the curvature
perturbations generated from this model. In the present case, the coefficients 
 $\mathcal{F}_S$ and $\mathcal{G}_S$ are given, from Eqs.~(\ref{pertFs}) and (\ref{pertGs}), by
\begin{align}
 \mathcal{F}_S &\simeq - \frac{2m \varphi^{2n+1}
  \dot{\varphi}^{2m+1}}{2^{m-1}HM_{Pl}^2M^{2n+4m}}, \nonumber \\   
 \mathcal{G}_S &\simeq - \frac{3m^2 \varphi^{2n+1}
  \dot{\varphi}^{2m+1}}{2^{m-1}HM_{Pl}^2M^{2n+4m}}.
\end{align}
This leads to the sound speed
\begin{equation}
c_s^2 := \frac{\mathcal{F}_S}{\mathcal{G}_S} \simeq \frac{2}{3m}.
\end{equation}
The curvature
perturbation in the present case is expressed in terms of $N$ as
\begin{align}
  \mathcal{P}_\zeta &= \frac{1}{8\pi^2 c_s \mathcal{F}_S}
  \left(\frac{H}{M_P}\right)^2 = - \frac{2^{m-5} H^3 M^{4m+2n}}{m \pi^2 c_s \varphi^{2n+1} \dot{\varphi}^{2m+1}} 
  \nonumber \\
  &= \pi^{-2} \left[
               2^{-\frac{39m+17n+13}{2}} 3^{\frac{7m+n-3}{2}} m^{\frac{3m+n-3}{2}}   
               \lambda^{m+n+1} \right. \nonumber \\
  & \qquad \left.   M_P^{-4(2m+n)} M^{4(2m+n)}
               \left( \frac{3m+n+1}{m}N + 2 \right)^{7m+n+1} 
             \right]^{\frac{1}{3m+n+1}}.
\label{Gpower}
\end{align}
By solving this equation for $M$, we obtain the expression for the model
parameter $M$:
\begin{align}
M = \left[
      \left(\mathcal{P}_\zeta \pi^2 \right)^{3m+n+1}
      2^{\frac{39m+17n+13}{2}} 3^{-\frac{7m+n-3}{2}} m^{-\frac{3m+n-3}{2}}
      \lambda^{-(m+n+1)} \left( \frac{3m+n+1}{m}N + 2 \right)^{-(7m+n+1)} 
    \right]^{\frac{1}{4(2m+n)}}M_P.
\end{align}
Substituting this back into Eq.~(\ref{varphiend}), we acquire the expression for the value of the Higgs field:
\begin{equation}
\varphi \simeq [\mathcal{P}_\zeta \pi^2]^\frac{1}{4}
               2^{\frac{17}{8}} 3^{-\frac18} m^{-\frac{1}{8}} 
               \lambda^{-\frac{1}{4}} 
               \left( \frac{3m+n+1}{m}N+2 \right)^{-\frac{1}{4}} M_{P}. \label{gphi}
\end{equation}

The spectral index can be calculated using the expression for the power
spectrum Eq.~(\ref{Gpower}):
\begin{align}
n_s -1 = \frac{d\ln \mathcal{P}_\zeta }{d\ln k}&= 3\frac{d\ln H}{d\ln k} - (2n+1)\frac{d\ln \varphi}{d\ln k} - (2m+1) \frac{d\ln \dot{\varphi}}{d\ln k} \notag\\ 
&=-3\epsilon {- \alpha} +(2m+1)\eta. %+(2n+1) \alpha +(2m+1)\eta.
\end{align}
Inserting the expressions for the slow-roll parameters Eq.~(\ref{eq:gslow})
gives
\begin{equation}
  n_s - 1 \simeq -\frac{7m+n+1}{N(3m+n+1)+2m},
\end{equation}
which agrees with the calculation in the main text.

The tensor power spectrum is of the same expression as before,
\begin{equation}
\mathcal{P}_t = \frac{8}{M_P^2}\left(\frac{H}{2\pi}\right)^2.
\end{equation}
Thus, the tensor-to-scalar ratio becomes
\begin{equation}
r:= \frac{\mathcal{P}_t}{\mathcal{P}_\zeta} 
 = 16c_s \mathcal{F}_S
 = - \frac{c_s {m}\varphi^{2n+1} \dot{\varphi}^{2m+1}}
          {2^{m-6}HM_P^2M^{4m+2n}}
 \simeq \frac{2^{\frac{15}{2}} 3^{-\frac32} m^{\frac12} }{N(3m+n+1)+2m}.
\end{equation}
The tensor-to-scalar ratio becomes smaller for larger $n$, which is the
region where the analysis here is valid.

\end{document}